\documentclass[preprint]{aastex}
\shortauthors{Xiang et al.}
\usepackage{multirow}
\usepackage{epsfig}
\usepackage{subfigure}
\usepackage{graphicx,times}
\input{epsf.sty}                        

\begin{document}
\title{Evolutionary Characteristics of the Solar Coronal Rotation: Revealed by a Century of Observations}
\author{N. B. Xiang\altaffilmark{1,2}, X. H. Zhao\altaffilmark{2}, S. W. Feng\altaffilmark{3}, C. Y. Shi\altaffilmark{2,4},  X. Y. Yu\altaffilmark{2,4}}
\affil{$^{1}$Yunnan Observatories, Chinese Academy of Sciences, Kunming 650216, China}
 \affil{$^{2}$State Key Laboratory of Solar Activity and Space Weather, National Space Science Center, Chinese Academy of Sciences, Beijing 100190, China; xhzhao@spaceweather.ac.cn}
 \affil{$^{3}$School of Space Science and Technology, Shandong University, Weihai 264209, People’s Republic of China}
\affil{$^{4}$University of Chinese Academy of Sciences, Beijing 100049, People’s Republic of China}


\begin{abstract}
The 10.7 cm radio flux time series from 1874 May 9 to 2017 June 30, the longest available observational proxy of the solar corona, is used to systematically investigate the multi-scale evolution of coronal rotation period lengths (RPLs) using wavelet analysis and linear regression. Throughout the time interval considered, the coronal rotation is characterized by a dominant synodic period of approximately 26.72 days, which varies in a complex manner with the advancing solar cycle. The relationship between the coronal RPLs and sunspot number (SN) changes over time: anticorrelation during 1874.9--1943.1, correlation during 1943.1--1958.2, anticorrelation during 1958.2--1987.4, and correlation during 1987.4--2016.9. Overall, RPLs show a negative correlation with SN, but their relationship is complex. This scenario is reflected in the phase dependence of the RPLs on the solar cycle. The variation of RPLs is complex, but a distinct pattern emerges: during weak activity phases, coronal rotation is slower, whereas during strong activity phases (approximately years 3 to 6 of the solar cycle), it becomes faster. Several statistically significant periods are detected in RPLs. Cross wavelet transform and wavelet coherence further confirm that the complex evolutionary pattern of RPLs fundamentally results from the combined modulation of two factors: the direct and dominant effect of the ubiquitous small-scale magnetic fields on the coronal rotation, and the modulation of the entire solar atmospheric rotation by the solar magnetic activity level.

\end{abstract}

\keywords{coronal rotation --- solar magnetic fields}

\section{Introduction}
The  rotation of the Sun was first observed in the early 17th century by Galileo using his pioneering optical telescope. Building upon these initial sunspot observations, Scheiner identified differential rotation in 1630, noting that sunspots near the equator rotated faster than those at higher latitudes. This phenomenon was later quantified by Carrington in 1863 through systematic sunspot observations (Carrington 1863; Patern{\`o} 2010; Obridko \& Badalyan 2020). Since that time, solar rotation has remained one of the most fundamental subjects in solar physics. Particularly in recent decades, with the advancements in ground-based and space-based observational capabilities, coupled with the accumulation of extensive data, a substantial number of studies have been conducted to thoroughly investigate rotation from the solar interior through to the outer coronal atmosphere (Howard 1984; Howe 2009; Kutsenko et al. 2023; L{\"o}{\ss}nitz et al. 2025).  However, solar rotation is widely recognized as a complex phenomenon, largely due to the significant variations in results obtained from diverse observational data and analytical techniques.

From the solar interior to the photosphere, the complexity of solar rotation stems from the interdependence between the solar dynamo and convective plasma. This interdependence leads to variations in solar rotation rate with respect to latitude, depth within the Sun, and the phase of the solar cycle (Babcock 1961; Thompson et al. 2003; Charbonneau 2010, 2020; Cameron et al. 2017; Hotta \& Kusano 2021). Thus, the study of solar rotation serves as the foundation for understanding this interdependence, and is consequently essential for understanding the dynamo mechanism. In the solar atmosphere, the rotation exhibits even more complex variations. This is due to the differing modulation effects of magnetic activities at various scales on the rotation, as well as changes in the plasma $\beta$ with altitude. These factors lead to different scenarios, such as the plasma controlling the rotation, or, under conditions of a strong frozen-in effect, the magnetic field controlling the plasma and thereby dominating the solar rotation (Gary 2001; Bellot Rubio \& Orozco Su{\'a}rez 2019; Li et al. 2019; Rao et al. 2024; Xiang et al. 2024).

Furthermore, studies of coronal rotation in the Sun's upper atmosphere encounter additional challenges. The corona is optically thin to extreme ultraviolet radiation and lacks persistent structural features that could serve as clear, long-lived tracers of the rotation in both space and time. At the same time, accurate direct measurements of the coronal magnetic field remain unavailable to date (Bhatt et al. 2017; Mancuso et al. 2020; Edwards et al. 2022; Xiang et al. 2024). Consequently, significant discrepancies exist among the findings of previous studies on coronal rotation, and a definitive consensus has yet to be reached (Vats et al. 2001; Braj{\v{s}}a et al. 2004; Mancuso \& Giordano 2011; Bhatt et al. 2017; Sharma et al. 2023; Kumar et al. 2024; Xiang et al. 2024). However, the complex rotation within the solar corona facilitates magnetic reconnection, thereby releasing the energy stored in the magnetic field, which results from the shearing and twisting motions driven by the differential rotation of the underlying photosphere. Understanding this process is a key to explaining the dynamics and heating of the corona (Zweibel \& Yamada 2009; Li et al. 2018; Rao et al. 2024). Simultaneously, large-scale magnetic structures in the corona, which may be anchored below the photosphere, can partially reflect the rotation of the solar interior. This connection provides valuable additional information for the  solar dynamo theory (Hiremath \& Hegde 2013; Hathaway 2015; Deng et al. 2020; Mancuso et al. 2020; Routh et al. 2024). Additionally, the rate of coronal rotation serves as a critical input for solar-wind models. Here, deviations of even about one degree per day or more can introduce significant systematic errors in forecasts that span longer than several days (Edwards et al. 2022).

Studies of coronal rotation primarily involve analyzing spectral features formed in the corona, such as the Fe {\sc xiv}  and Fe {\sc x}  emission lines (Altrock 2003; Badalyan \& S{\'y}kora 2006; Badalyan \& Obridko 2017), as well as soft X-ray (Weber et al. 1999; Weber \& Sturrock 2002; Chandra et al. 2010), ultraviolet rays (Mancuso \& Giordano 2011; Sharma et al. 2020; Routh et al. 2024), and radio emissions (Vats et al. 2001; Singh et al. 2021; Xiang et al. 2024). Coronal holes and coronal bright points are also commonly used as tracers to study coronal rotation (Navarro-Peralta \& Sanchez-Ibarra 1994; Braj{\v{s}}a et al. 2004; Hiremath \& Hegde 2013; Oghrapishvili et al. 2018; Shokri et al. 2024; Sudar et al. 2024). Overall, the corona rotates slightly faster than the underlying photosphere (Livingston 1969; Li et al. 2019; Xiang et al. 2023; Routh et al. 2024). The corona also exhibits differential rotation, but with a weaker latitudinal gradient compared to the photosphere. At low latitudes, the differential rotation profile and rate of the corona are similar to those of sunspots in the photosphere. However, at middle to high latitudes, coronal rotation becomes more rigid and faster (Lewis et al. 1999; Giordano \& Mancuso 2008; Chandra et al. 2009; Mancuso \& Giordano 2011; Dorotovi{\v{c}} \& Rybansk{\'y} 2019; Mancuso et al. 2020; Sharma et al. 2023; Zhao et al. 2026).

It is important to note that these findings on coronal rotation vary depending on the tracers or flux modulation methods used, the time intervals analyzed, and the phase of the solar cycle. Thus, some  studies  especially explored the relationship between coronal rotation and the solar cycle, revealing that both the latitudinal gradient of coronal differential rotation and its rotation rate are correlated with the solar cycle (Altrock 2003; Chandra et al. 2010; Badalyan \& Obridko 2018; Edwards et al. 2022; Routh et al. 2024). For instance, analyses of the green coronal line enabled Sime et al. (1989) and Altrock (2003) to identify a pattern of weak differential rotation that varies with the solar cycle. Their results describe a transition between predominantly rigid rotation near solar minimum and more differential rotation near solar maximum. In contrast, Badalyan and Obridko (2018) reported that coronal rotation exhibits the most pronounced differential characteristics during solar minimum. Edwards et al. (2022) found that  the solar corona exhibited a specific rotational pattern: its rotation rate at almost all latitudes abruptly slowed down as solar activity entered an ascending phase, and then returned to a faster state near the subsequent solar minimum. Commonly, these studies typically used data spanning less than two solar cycles.

On even longer timescales spanning decades, the temporal evolution of coronal rotation is primarily studied from a global perspective using the flux modulation method, focusing on emissions originating from the corona (Chandra \& Vats 2011; Xie et al. 2017; Singh et al. 2021, Xiang et al. 2024). For instance, by examining the 10.7 cm radio flux data spanning solar cycles 19 to 22, Mouradian et al. (2002) investigated the relationship between coronal rotation and solar activity. Their analysis revealed that the synodic period decreases from about 30 days at solar minima to 25.4 days at solar maxima. Based on an autocorrelation method, Chandra and Vats (2011)  determined the annual rotation periods from daily 10.7 cm solar radio flux data between 1947 and 2009, and further investigated the long-term variation of rotation period lengths (RPLs) of the solar corona. Their analysis revealed a 22-year periodicity in the RPLs, which is associated with the solar magnetic reversal cycle. Li et al. (2012) later applied empirical mode decomposition to analyze both the yearly coronal rotation periods derived by Chandra \& Vats (2011) and the corresponding annual mean 10.7 cm solar radio flux. Their results indicated a slight long-term decrease in the coronal RPLs, while no statistically significant Schwabe cycle was identified in the variation of the RPLs. Xie et al. (2017) revisited the temporal variation of coronal rotation using daily 10.7 cm solar radio flux data from 1947 to 2014. Their analysis also revealed a long-term decreasing trend in the RPLs over the time interval. Moreover, several significant periodicities, ranging from about 3 to 11 years, were detected in the variation of the coronal RPLs. The authors proposed that the solar cycle variation in RPLs is linked to the Schwabe solar cycle. Similarly, Deng et al. (2020) utilized a revised coronal index covering the period 1939-2019 to investigate the long-term evolution of coronal rotation, and  their analysis arrived at conclusions largely consistent with those reported earlier by Xie et al. (2017). This correlation between the variations of coronal RPLs and the solar activity cycle stands in contrast to the earlier findings of Chandra \& Vats (2011) and Li et al. (2012). So far, most studies on the long-term evolution of coronal rotation have relied on the same or similar datasets and have examined nearly identical time intervals. Nevertheless, their findings are inconsistent and even contradictory in some aspects. To date, the variation pattern of coronal rotation over longer timescales, as well as the dependence of its RPLs on the phase of the solar activity cycle, remains unclear. Furthermore, previous research has seldom addressed the physical mechanisms underlying the multi-scale variations in the RPLs of the solar corona. Therefore, further studying the long-term evolution of coronal rotation and how its variation depends on the solar cycle phase is both highly necessary and of great significance.

The 10.7 cm radio flux is reconstructed using an empirical model developed by Yeo, Solanki, and Krivova (2020), which extends the approach of Preminger \& Walton (2005, 2006a, 2006b, 2007). The model relates sunspot indices (e.g., total sunspot area) to the 10.7 cm flux by convolving the sunspot data with a finite impulse response filter that accounts for active region evolution and a possible nonlinear relationship. The reconstructed data offer extended temporal coverage (back to earlier sunspot records), better physical consistency with solar magnetic behavior, and improved accuracy in reproducing observed variability compared to earlier models. The 10.7 cm radio flux data reconstructed by Yeo et al. (2020) spans from 1874 to 2017, covering 143 years, which stands as the longest time series currently available for characterizing solar coronal activity. Furthermore, the analysis of fractal dimensions in the solar radio flux time series reveals that emissions at 2800 MHz (corresponding to a 10.7 cm wavelength) exhibit the lowest degree of randomness, making it the most suitable frequency for studying coronal rotation (Vats et al. 1997; Chandra \& Vats 2011; Bhatt et al. 2018; Xiang et al. 2024). Consequently, in this study, we utilize the longest available time series of 10.7 cm radio flux to analyze the multi-scale temporal evolution of coronal rotation, ranging from annual to centennial timescales, and further uncover the underlying physical mechanisms.

\section{Data and Methods}
\subsection{Data}
Based on the empirical modeling approach, Yeo et al. (2020) reconstructed  the 10.7 cm solar radio flux time series  to extend its reliable representation of the coronal activity further back in time. The reconstruction utilizes the group and international sunspot numbers (SN) or total sunspot area as input proxies.  The core methodology, constructed by Yeo, Solanki, and Krivova, is known as the YSK model, which represents an improvement and extension of the linear transformation framework proposed  by Preminger \& Walton (2005, 2006a, b, 2007). It operates by convolving a sunspot index with a model finite impulse response filter, which mathematically encodes the temporal response of facular and network magnetic flux to the emergence and decay of sunspot-bearing active regions (Yeo et al., 2020). A key refinement in the YSK model is the introduction of a non-linear power-law relationship between the sunspot index and the target radio flux, better capturing the physics of chromospheric emission and magnetic flux relationships compared to a purely linear assumption. The reconstructed dataset offers significant advantages. Firstly, it provides a physically consistent and continuous record of a facular proxy, validated against direct measurements of chromospheric indices and photospheric magnetic flux over the satellite era (Yeo et al., 2020). Secondly, the model successfully replicates most observed variability (up to $\sim$94\%), outperforming earlier empirical models like those of Tapping \& Vald{\'e}s(2011) and the original model by Preminger \& Walton (2005, 2006a, b, 2007). Crucially, the analysis using this improved model found no robust evidence for a secular change in the relationship between the radio flux and sunspot indices over recent decades, contrary to some previous studies whose simpler methodologies may have introduced artefacts (Yeo et al., 2020). Thus, the reconstructed series constitutes a more stable and reliable tool for investigating long-term solar variability.

\begin{figure*}[!htbp]
\begin{center}
\includegraphics[width=1.0\textwidth]{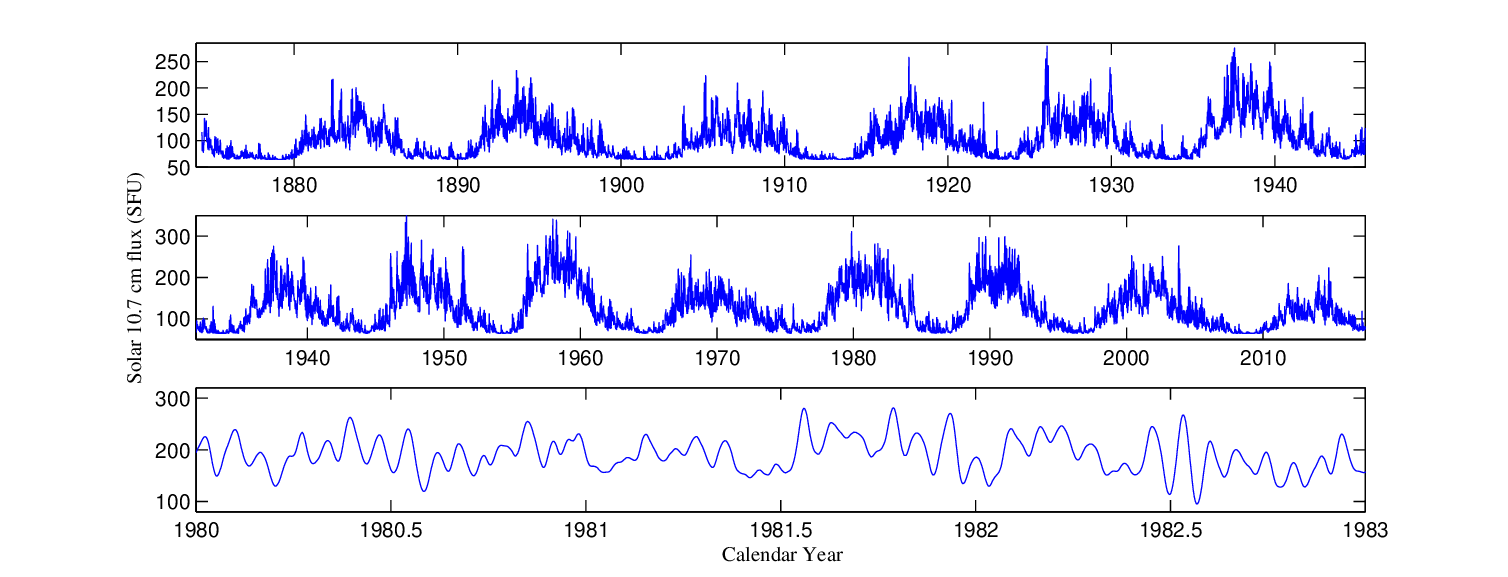}
\end{center}
\caption{Top and middle panels: the 10.7 cm solar radio flux time series from 1874 May 9 to 2017 June 30, arranged in chronological order. Bottom panel: a randomly selected three-year segment (1980.0-1983.0) extracted from the time series.}
\end{figure*}

The 10.7 cm solar radio flux time series reconstructed by Yeo et al. (2020) is publicly available from the SATIRE model reconstruction of total and spectral solar irradiance (Solar Variability And Climate Group, 2017).\footnote{\url{http://www2.mps.mpg.de/projects/sun-climate/data.html}} This reconstructed dataset spans from 1874 May 9 to 2017 June 30, covering the declining phase of solar cycle 11 through the late phase of solar cycle 24. The time series comprises a total of 52,283 days, of which 51,077 days have valid records, with only 1,206 days (approximately 2.31\%) missing, rendering it highly suitable for this study. Therefore, we use this reconstructed 10.7 cm solar radio flux time series to investigate the multi-scale  temporal  evolution of the coronal rotation period over a centennial timescale. The reconstructed 10.7 cm solar radio flux time series is presented in chronological order in the top and middle panels of Figure 1. Additionally, to examine whether the 10.7 cm radio flux exhibits rotational period signals, a randomly selected three-year segment from 1980.0 to 1983.0 is extracted from the time series and displayed in the bottom panel of the figure.

As shown in Figure 1, the x-axis denotes the calendar year, while the y-axis represents the 10.7 cm solar radio flux, expressed in solar flux units (SFU; \(1\,\mathrm{SFU} = 10^{-22}\,\mathrm{W\,m^{-2}\,Hz^{-1}}\)). The 10.7 cm radio flux time series exhibits a clear ~11-year solar activity cycle, with higher values and enhanced variability during solar maxima. This pattern stems from the close relationship between the 10.7 cm radio flux and photospheric magnetic activity. In the bottom panel, the  modulation due to the coronal rotation is evident, suggesting that the 10.7 cm radio data is well suited for investigating the coronal rotation period and its temporal evolution.

\subsection{Methods}
Wavelet analysis, as a time-frequency analysis tool, decomposes one-dimensional signals into two-dimensional representations, yielding localized power spectra that capture both temporal and frequency characteristics. This approach enables the identification of dominant periodicities across varying timescales while preserving information about when these oscillations occur (Torrence \& Compo 1998; Grinsted et al. 2004). Consequently, wavelet analysis has been widely used to extract multi-scale periodic patterns from solar activity time series and to track their temporal evolution, with recent applications across various solar flux measurements to investigate the rotation of the solar atmosphere and its variations (Gurgenashvili et al. 2016; Ashfield et al. 2026; Xiang et al. 2024, 2026). In this study, we apply this technique to the 10.7 cm solar radio flux time series to specifically investigate its rotation period and the temporal evolution of its rotational characteristics. We employ the continuous wavelet transform (CWT) with the Morlet wavelet as the mother function, selected for its strong capability in feature extraction and its advantageous time-frequency localization (Torrence \& Compo 1998; Grinsted et al. 2004; Xiang et al. 2024). Through systematic parameter testing, we determined that a dimensionless frequency of $\omega_{0}=12$ provides an optimal compromise between time and frequency resolution. Statistical significance of the wavelet power spectra is assessed under the assumption of a red-noise background, with periodicities identified at a stringent 99\% confidence level to ensure the reliability of the results.

Based on the CWT of two time series, cross wavelet transform (XWT) and wavelet coherence (WTC) can be constructed. The XWT and WTC are both powerful tools for analyzing the relationship between two non-stationary time series in the time-frequency domain. They are like ``twins in time-frequency analysis", possessing both close connections and a distinct division of labor. Simply put, XWT excels at identifying regions where both signals have strong power, while WTC is adept at finding areas where their patterns of change are correlated, even if their power is weak (Torrence \& Compo 1998; Grinsted et al. 2004; Xu et al. 2017).

For two individual time series $X_{n}, n=1, 2, 3, ..., N$ and $Y_{n}, n=1, 2, 3, ..., N$, the XWT is constructed from the CWTs of the two series, denoted as  $W^{X}$ and $W^{Y}$, and is defined as $W^{XY}=W^{X}W^{Y*}$, where the asterisk indicates complex conjugation. The magnitude of the XWT,  $|W^{XY}|$,  quantifies the cross wavelet power, representing regions of high common power in the time-frequency space. The complex argument $arg(W^{XY} )$ gives the local relative phase between the two series at each time and scale within this time-frequency domain, thereby revealing their phase synchronization or delay over a broad range of timescales, from months to several decades in this study. However, phase estimates at the longest periods should be interpreted with caution, as they are limited by the finite length of the time series and the cone of influence (Torrence \& Compo 1998; Grinsted et al. 2004; Xiang \& Kong 2015). Statistical significance of the XWT can be assessed against a red noise background by considering the theoretical distribution of the cross wavelet power, which is derived from the product of two $\chi^{2}$ distributions associated with the background spectra of the individual series (Jenkins \& Watts 1969; Torrence \& Compo 1998; Grinsted et al. 2004). This methodology enables the detection of common power and phase relationships even when individual series exhibit low power.

The WTC constitutes another valuable technique that quantifies the degree of coherence of the XWT within the time-frequency domain, enabling the identification of significant coherence even in regions where the common power between the two series is low. It can be interpreted as a localized correlation measure between two time series in the time-frequency space, thereby offering a robust approach for examining hypothesized linkages between them (see Grinsted et al., 2004 for details). Moreover, given that the cross wavelet spectrum is unsuitable for significance testing of interrelations between two processes (Maraun \& Kurths 2004; Xiang \& Kong 2015), the application of WTC becomes essential. The WTC between two time series,$X_{n}$ and $Y_{n}$, can be  expressed as (Grinsted et al., 2004; Li et al. 2009)
\begin{equation}
  R_{n}^{2}(s)=\frac{|S(s^{-1}W_{n}^{XY}(s))|^{2}}{S(s^{-1}|W_{n}^{X}(s)|^{2})\cdot S(s^{-1}|W_{n}^{Y}(s)|^{2})}
\end{equation}
where $S$ represents a smoothing operator. Given its structural similarity to the classical correlation coefficient, the WTC can be conceptualized as a localized correlation coefficient in the time-frequency domain. The smoothing operator $S$ is given by
\begin{equation}
  S(W)=S_{scale}(S_{time}(W_{n}(s)))
\end{equation}
where $S_{scale}$ and $S_{time}$  denote smoothing operations performed along the wavelet scale axis and in the time domain, respectively. Given that the Morlet wavelet is chosen for this study, an appropriate smoothing operator can be defined by the following functions (Grinsted et al., 2004; Xiang \& Kong 2015):
\begin{equation}
S_{time}(W)|_{s}=(W_{n}(s)*c_{1}^\frac{-t^{2}}{2s^{2}})|_{s}
\end{equation}
\begin{equation}
S_{scale}(W)|_{n}=(W_{n}(s)*c_{2}\prod(0.6s))|_{n}
\end{equation}
where $c_{1}$ \& $c_{2}$ represent normalization constants, and $\prod$ denotes the rectangle function. Following the empirically determined scale decorrelation length for the Morlet wavelet (Torrence \& Compo 1998; Grinsted et al. 2004), the factor is set to 0.6. In practice, both convolutions are performed discretely, requiring the normalization coefficients to be determined numerically. Furthermore, the statistical significance of the WTC is assessed using Monte Carlo methods (Grinsted et al. 2004; Xiang \& Kong 2015).
In this study, the XWT and WTC are employed to investigate, on centennial and decadal timescales, the regions of high common power and the localized correlation measure between coronal rotation and solar magnetic activity in the time-frequency space, with the aim of exploring the causes of multi-timescale variations in coronal rotation.

\section{Multi-scale Temporal Evolution of the Coronal Rotation Period}
\subsection{Rotation Period of the corona}
The data gaps  present in the 10.7 cm solar radio flux time series during the time interval we considered were addressed through linear interpolation. A total of 1206 missing data points were distributed randomly throughout the 52,283--day record, and the longest gaps generally did not exceed 5 days, considerably shorter than the 27--day solar rotation period. In order to assess the potential impact of data gaps on our results, we also applied two alternative gap--filling methods: (i) cubic spline interpolation, and (ii)  replacement with the average of their neighboring values of the 10.7 cm solar radio flux. The rotation periods derived from all three interpolation schemes were nearly identical, confirming that our findings are robust and insensitive to the choice of interpolation method. Consequently, given that the choice of interpolation method has a negligible effect on the derived rotation periods, linear interpolation was adopted for the subsequent analysis. The CWT was then applied to the composite 10.7 cm flux time series to examine its periodic behavior. The resulting wavelet power spectra are displayed chronologically from top to bottom in Figure 2, with black contours indicating the 99\% confidence level. Since the primary  purpose is to investigate the rotation period and its temporal evolution, the wavelet analysis focuses on the period range spanning from 16 days to 45 days. The global spectrum for the 10.7 cm solar radio flux is depicted as a solid line in the right column of the bottom panel in Figure 2, together with its corresponding 99\% confidence level represented by a dashed line.

\begin{figure*}[h!]
\begin{center}
\includegraphics[width=1.0\textwidth]{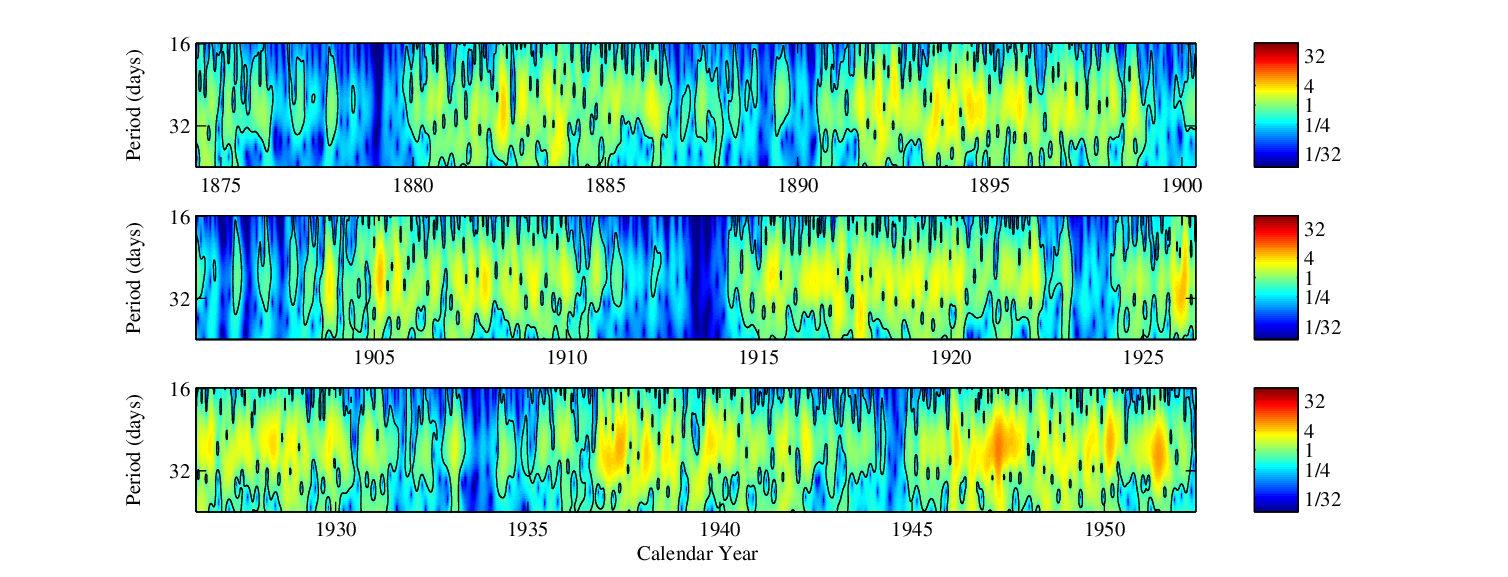}
\includegraphics[width=1.0\textwidth]{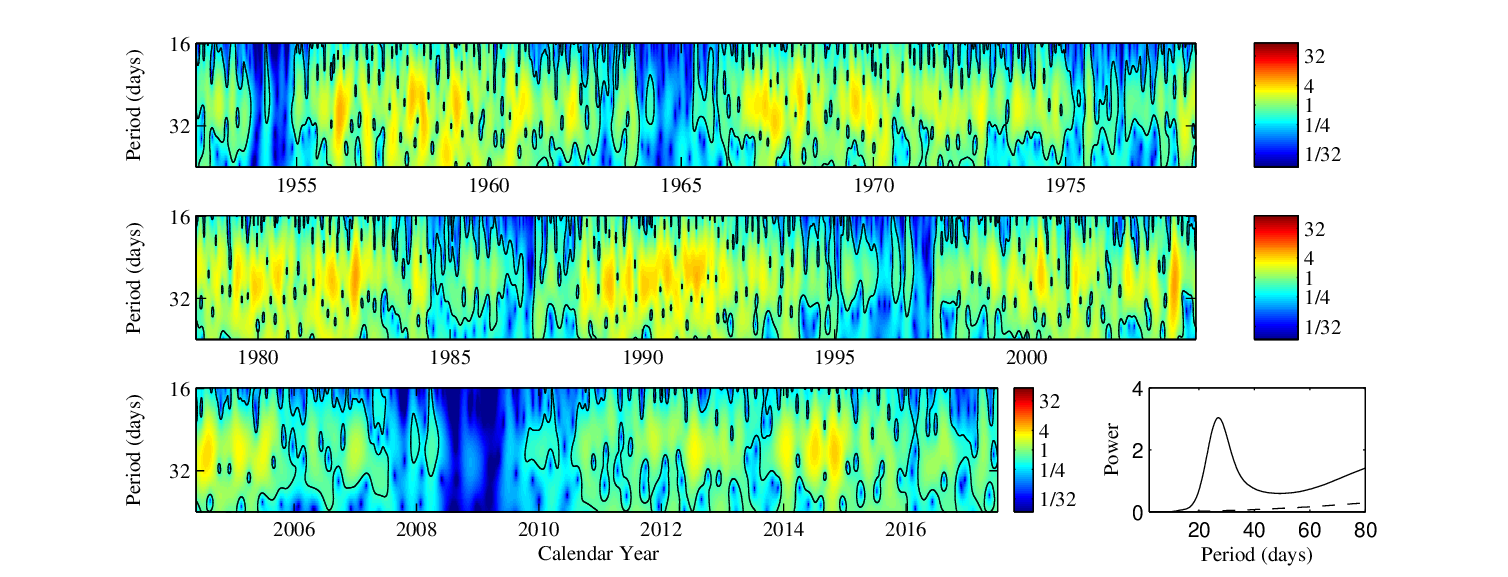}
\end{center}
\caption{Continuous wavelet power spectra of the 10.7 cm solar radio flux time series spanning  from 1874 May 9 to 2017 June 30, organized in chronological order from top to bottom. In each panel, black contours represent the 99\% confidence level. To better illustrate the rotation period and its evolution, the spectra are shown only for timescales between 16 and 45 days. Right column of the bottom panel: Global wavelet power spectrum (solid line) and the corresponding 99\% confidence level (dashed line) for the 10.7 cm solar radio flux time series over the entire study interval.}
\end{figure*}

As shown in Figure 2, the continuous wavelet power spectra reveal that the approximately 27-day rotation period of the 10.7 cm solar radio flux is statistically significant throughout nearly the entire study interval. However, the regions exhibiting statistical significance at the solar rotation timescale clearly display localized oscillatory features, indicating that the rotation period of the 10.7 cm flux undergoes temporal evolution as the solar cycle progresses. In the global wavelet spectrum, a narrow and sharp peak enables precise determination of the dominant synodic rotation period of the corona, which is found to be 26.72 days.

\subsection{Long-term  trends of the Rotation Period Lengths of the corona}
The long-term variation in the rotation rate of different solar atmospheric layers has become a significant issue, as it may reveal the coupling relationship between solar
atmospheric rotation and solar magnetic activity on extended timescales (from decades to longer) (Chandra \& Vats 2011; Zhang et al. 2023; Xiang et al. 2026). Wavelet analysis allows the extraction of localized oscillatory features of the rotation period, yielding the signal of its variation over the progression of solar cycles. Furthermore, as the 10.7 cm radio flux time series analyzed in this study spans more than 100 years, it enables an investigation into the long-term trend of the coronal rotation period. As shown in Figure 2, the daily local wavelet power spectra of the 10.7 cm radio flux time series exhibit distinct peaks at the rotational timescales, enabling the determination of the rotation period throughout the entire study interval. This method has been widely employed in previous studies, but the specific range of timescales selected to represent the rotation period varies across different layers of the solar atmosphere. For instance, Xu et al. (2020) used a range of 22-33 days to study the temporal evolution of chromospheric rotation, while Xie et al. (2017) adopted a range of 25-31 days for coronal rotation. In comparison, Chandra and Vats (2011) analysed the 10.7 cm solar radio flux time series over the period 1947-2009 and reported a sidereal rotation period range of 19.0-29.5 days, which corresponds to a synodic rotation period range of approximately 20.1-32.1 days. On the basis of both the local characteristics of the wavelet power spectrum of the 10.7 cm radio flux time series and the findings of Chandra and Vats (2011), we therefore selected a period range of 20-32 days to investigate the coronal rotation. To suppress short-term fluctuations, the daily rotation periods derived from the 10.7 cm radio flux were smoothed using a 1 yr sliding average. The resulting time series, depicted by the solid blue line in the middle panel of Figure 3, illustrates how the coronal RPLs vary in response to solar activity. The figure also reveals that the coronal RPLs exhibit complex quasi-periodic variations on the timescales of several years to decades, along with distinct trends of lengthening and shortening across different time intervals.

\begin{figure*}[!htbp]
\begin{center}
\includegraphics[width=1.0\textwidth]{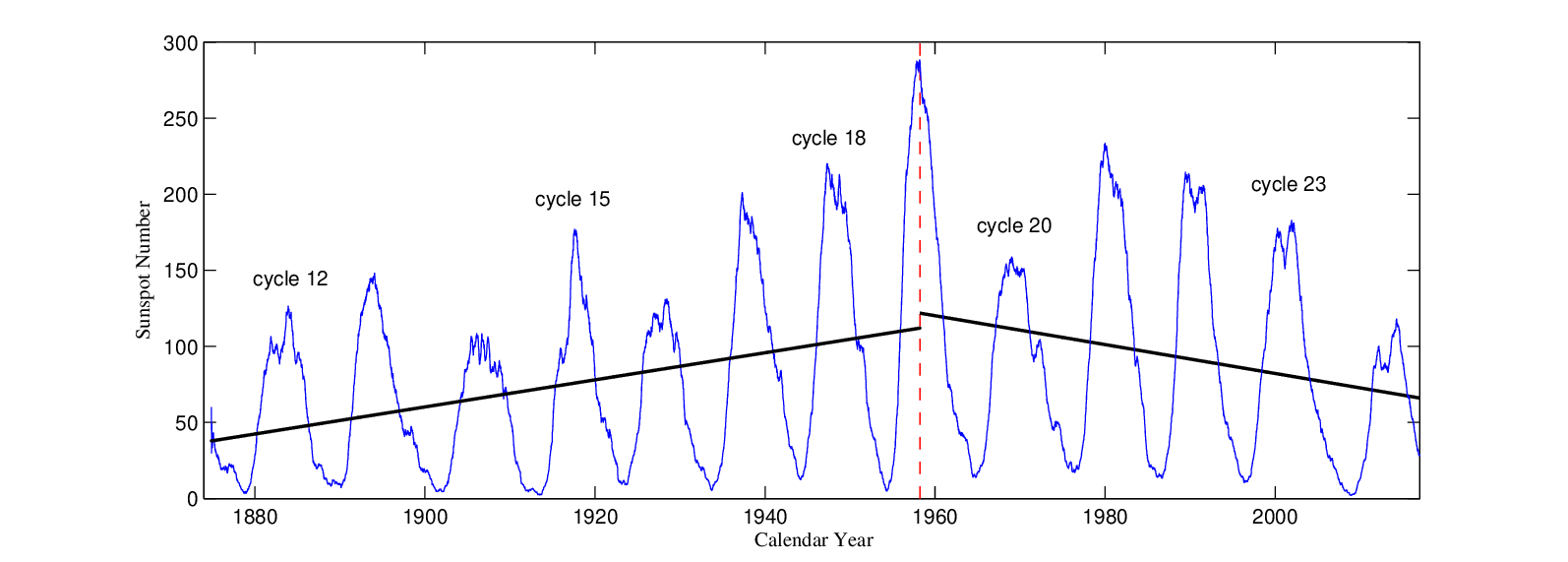}
\includegraphics[width=1.0\textwidth]{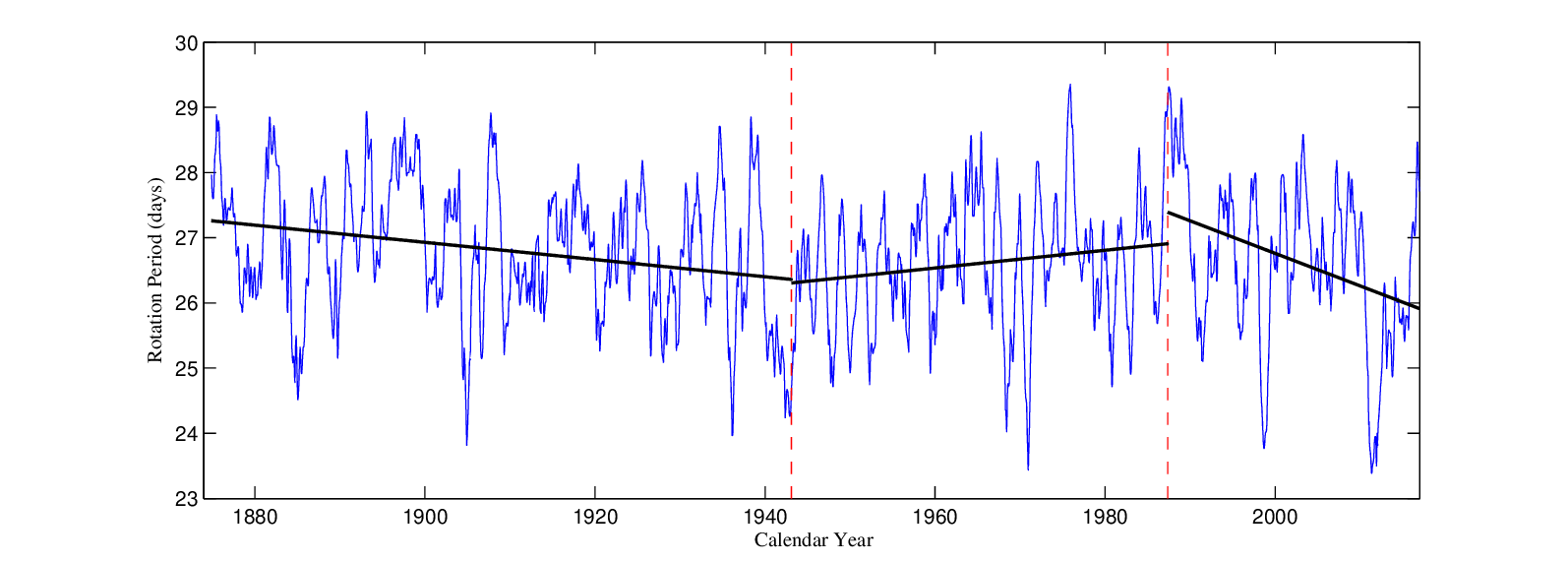}
\includegraphics[width=1.0\textwidth]{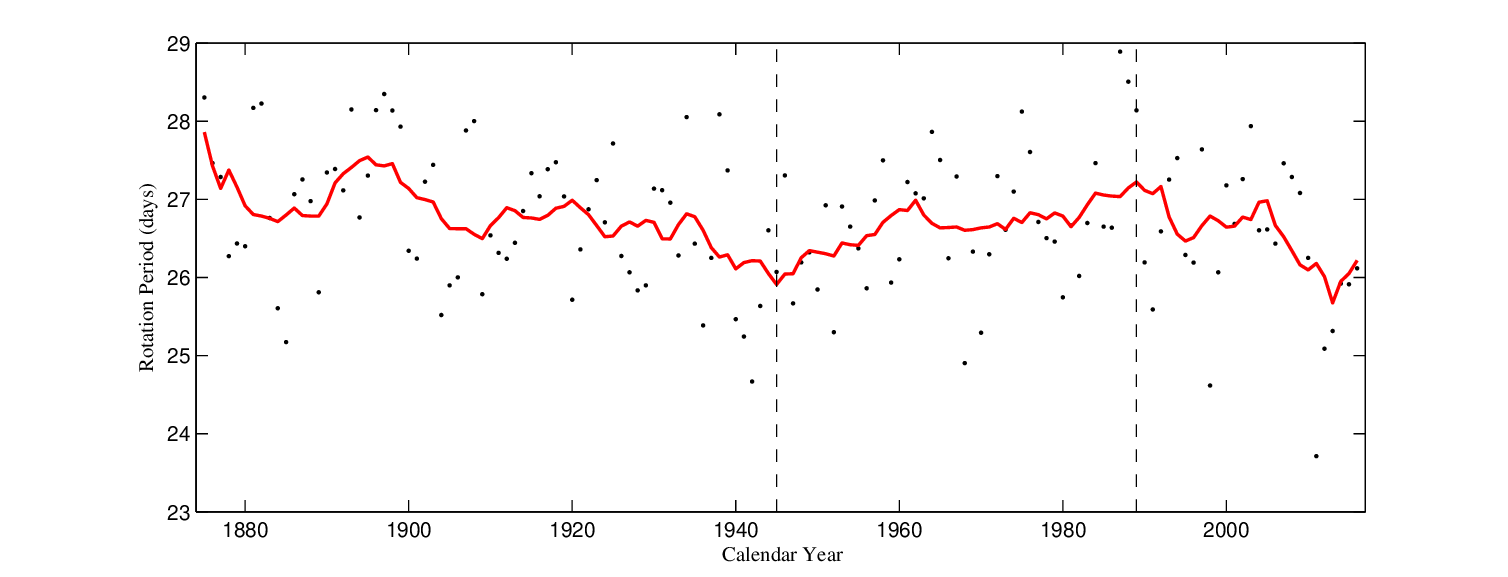}
\end{center}
\caption{Top panel: the 1--year smoothed SN from 1874 May 9 to 2017 June 30  (solid blue line) and  the two linear regression segments (solid black lines). The red dashed vertical line marks the breakpoint at 1958.229.  Middle panel: temporal variation of the coronal rotation period during the  time interval considered (solid blue line) and  the corresponding three linear regression segments (solid black lines). The two red dashed vertical lines indicate the breakpoints at 1943.103 and 1987.374. Bottom panel: the yearly synodic rotation periods of the 10.7 cm solar radio flux from 1875 to 2016 (black dots) and their 11-year running mean (solid red line). The two black dashed vertical lines mark the turning points at 1945 and 1989.}
\end{figure*}

Based on observational data from recent decades, nearly all studies have simply employed linear regression analysis, revealing a gradual decrease in the  RPLs from the lower photosphere to the upper corona. The underlying reason may be that the gradual weakening of the solar magnetic field in recent decades has resulted in a progressive acceleration of rotation across the entire solar atmosphere (Heristchi \& Mouradian 2009; Li et al. 2012; Xie et al. 2017; Zhang et al. 2023). In contrast, Xiang et al. (2026) analyzed the composite plage area time series spanning 117 years and found that the long-term trend in the chromospheric rotation cannot be characterized by a linear function but instead requires a quadratic polynomial, exhibiting an initial deceleration followed by a subsequent acceleration. Similar to the aforementioned study, the 10.7 cm solar radio flux time series analyzed here spans a longer time interval (143 years) and exhibits a complex pattern of rotational period evolution, rather than a monotonic trend. Therefore, we further investigate the long-term variation in coronal RPLs in relation to changes in solar magnetic activity.

It is known that the SN is widely used as an indicator of the level of  solar magnetic activity (Solanki 2003; Hathaway 2015). The daily sunspot numbers (version 2.0; Clette et al. 2014; Clette \& Lef{\`e}vre 2016) from 1874 to 2017 used in this study were provided by the Sunspot Index and Long--term Solar Observations (SILSO) World Data Center, Royal Observatory of Belgium, Brussels (SILSO World Data Center, 1874--2017). To suppress short-term fluctuations, the time series was processed with a 1 yr sliding window, and the resulting curve is shown as the solid blue line in the top panel of Figure 3. As illustrated in the figure, although the smoothed SN still exhibits complex variation patterns, it can be summarized as two distinct trends: from solar cycle 12 to 19, the SN displays an increasing trend (despite relatively low values in cycles 14 and 16), while from cycle 19 to 24, it shows a gradual decreasing trend (similarly, despite relatively low values in cycle 20). To identify a statistically significant breakpoint in the SN series, we employed the  breakpoint detection procedure of Bai and Perron (2003). The analysis was performed on the full time series consisting of 52,283 data values. To quantify the uncertainty of the estimated break date, we implemented a bootstrap resampling approach with 200 replications (40,000 resamples each). This yielded a breakpoint at 1958.229 years with a 95\% confidence interval of [1958.116, 1958.426] years. This breakpoint is marked by the red dashed vertical line in the top panel of Figure 3. Then, we can employ a piecewise function consisting of two simple linear regressions to characterize the long-term trend of the SN.  For the first segment of the piecewise function, linear regression analysis was performed over the years from 1874.353 to 1958.229, yielding a slope of 0.8901$\pm$0.0257. For the second segment, covering years of 1958.229 to 2017.496, the resulting slope is -0.9489$\pm$0.0511. The corresponding t-statistic analysis indicates that the fitted results, which represent the long-term trend of the SN, are statistically significant. The fitted equations are given by:
\begin{equation}
Y_{SN}=
\left\{
	\begin{array}{lr}
	0.8901*t - 1631, & 1874.353 < t \leq 1958.229, \\
    -0.9489*t + 1980,  & 1958.229 < t \leq 2017.496,
	\end{array}
\right.
\end{equation}
where $Y_{SN}$ denotes  the fitted SN and $t$ represents  the time in years. The fitting results, depicted as  solid black lines in the top panel of Figure 3, clearly illustrate the long-term trend of the SN: a gradual increase from the initial year of 1874.353 to 1958.229, followed by a gradual decrease from 1958.229 to  the concluding year of 2017.496.

Similar to the long-term trend identified in the SN, the coronal RPLs also exhibit  distinct linear trends. To identify statistically significant breakpoints in the coronal RPL series, we applied the same breakpoint detection procedure of Bai and Perron (2003). However, as shown in the local wavelet power spectrum (Figure 2), the values  at both ends of the time series may be unreliable due to boundary effects and thus do not accurately reflect the true coronal RPLs. We therefore removed 200 data points from each end of the series. The final analysis was performed on the remaining 51,883 data points, covering the interval from 1874.901 to 2016.947 years. To quantify the uncertainty of the estimated break dates, we implemented a bootstrap resampling approach with 200 replications (40,000 resamples each), all of which converged successfully. This yielded two breakpoints: one at 1943.103 years, with the 95\%  confidence interval of [1942.270, 1943.714] years; and another at 1987.374 years, with the 95\%  confidence interval of [1986.667, 1987.905] years.
These two breakpoints are indicated by red dashed vertical lines in the middle panel of Figure 3. Therefore, a piecewise function consisting of three simple linear regression segments was used to fit the long-term variation of the coronal RPLs.  The three fitting intervals of the piecewise function are 1874.901-1943.103, 1943.103-1987.374, and 1987.374-2016.947, respectively. The slopes obtained from linear regression for these three intervals are -0.0132$\pm$0.0006, 0.0136$\pm$0.0011, and -0.0499$\pm$0.0025, respectively. The t-test analysis indicates that these fitting results, which reflect the long-term trend of coronal RPLs, are statistically significant. The fitting equations are as follows:
\begin{equation}
Y_{RPLs}=
\left\{
	\begin{array}{lr}
	-0.0132*t + 51.99, &  1874.901 < t \leq 1943.103, \\
    0.0136*t - 0.1204,  &  1943.103 < t \leq 1987.374,  \\
    -0.0499*t + 126.6, &  1987.374 <t \leq 2016.947,
	\end{array}
\right.
\end{equation}
where $Y_{RPLs}$ is the fitted coronal RPLs in days, and $t$ denotes the time in years. The fitting results are presented in the middle panel of Figure 3, where the three black solid lines represent the linear fits for the respective time intervals. As can be observed, over the time interval considered, the coronal RPLs exhibit three distinct long-term trends: an initial decrease, followed by an increase, and then a subsequent decrease.

Combining the top and middle panels of Figure 3, the fitting results for the long-term trends of the SN  and coronal RPLs reveal that, over the 141-year analysis time interval (the total time span is about 143 years, with 200 data points removed from both the beginning and the end), the trends of the two quantities are opposite for approximately 97 years, with a correlation coefficient of -0.783 between the fitted results, indicating a strong negative correlation. However, for the remaining 44 years, they exhibit the same trend, suggesting a complex relationship between them. To further verify the long-term trend of coronal RPLs, we employed the CWT again to determine the yearly synodic rotation period of the 10.7 cm solar radio flux from 1875 to 2016. The years 1874 and 2017 were excluded due to the availability of only approximately half a year of data in each. The resulting yearly rotation periods are shown as black dots in the bottom panel of Figure 3. Following the methods of Chandra \& Vats (2011) and Xie et al. (2017), we applied an 11--year running mean to these yearly rotation periods, with the smoothed results depicted by the red solid line in the bottom panel of Figure 3. The long-term trend of  the smoothed curve appears to exhibit two turning points, at 1945 and 1989, respectively, which are marked by black dashed vertical lines in the bottom panel of Figure 3. The results indicate that from 1875 to around 1945 (breakpoint at 1943.103 in the middle panel, slightly earlier than 1945), the coronal RPLs shortened gradually, indicating progressively faster rotation; from around 1945 to 1989 (breakpoint at 1987.374 in the middle panel, slightly earlier than 1989), the RPLs lengthened gradually, indicating progressively slower rotation; and from 1989 to the end of the study period, the RPLs shortened again, indicating progressively faster rotation. Over the entire 141--year analysis interval, the 11--year smoothed coronal RPLs exhibited a trend opposite to that of the SN for a total of 101 years, a duration slightly longer than that obtained from the linear fit in the middle panel of Figure 3, thereby indicating that the two methods yield consistent results. Overall, the RPLs are negatively correlated with the solar magnetic activity level, i.e., stronger magnetic activity corresponds to shorter coronal RPLs and faster rotation. Furthermore, the long-term variation of the coronal RPLs is not solely governed by SN; other factors also modulate their long--term evolution.

\subsection{Dependence of Coronal Rotation on the Solar Cycle Phase}
The mean rotation period for a given solar cycle phase is obtained by averaging the rotation periods at that phase across all cycles within the considered time interval. This method, which has been employed in numerous studies on solar atmospheric rotation, effectively reveals the dependence of the RPLs on the solar cycle phase (Braj{\v{s}}a et al. 2006; Xie et al. 2017; Zhang et al. 2023). Thus, based on the results shown in the middle panel of Figure 3 (solid blue line), the dependence of the coronal RPLs on the solar cycle phase can be obtained, where the phase is defined relative to the nearest preceding sunspot minimum. The results are displayed in Figure 4. As it shows, the relationship between the coronal RPLs and the solar cycle phase is complex. Nevertheless, the overall evolution of the RPLs exhibits three distinct stages. During the first stage, from the beginning of the solar cycle to approximately year 3, the RPLs gradually decrease, indicating progressively faster rotation. During the second stage, between approximately years 3 and 6, the RPLs remain at relatively low values, corresponding to persistently faster rotation. During the third stage, from approximately year 6 to the end of the solar cycle, the RPLs first increase between approximately years 6 and 7, then gradually decrease after year 7, and finally rise again toward the end of the solar cycle, indicating a complex temporal evolution. This third-stage variation suggests that the rotation first slows down, then speeds up, and slows down again near the end of the solar cycle, rather than becoming progressively slower. Despite this complexity, the rotation at the end of the solar cycle remains slower than during the solar activity maximum. Taken together, these results indicate that the coronal rotation is systematically slower during the years around solar minimum than during the years around solar maximum, reflecting a negative correlation between the coronal RPLs and sunspot activity, which is consistent with the findings of Xie et al. (2017). In addition, Badalyan \& Obridko (2017) used coronal green-line brightness data to analyze the 22-year cyclic variations in the parameters of differential rotation of the solar corona in odd and even activity cycles. They found that the equatorial rotation rate reaches a maximum near the even-odd cycle minimum and a minimum near the odd-even minimum, while the differential gradient remains more pronounced during the descending branch of even cycles and reaches its maximum later than in odd cycles. In this study, however, the coronal RPLs are examined as latitudinal averages and therefore cannot directly resolve the latitudinal differential gradient. Nevertheless, the complex non-monotonic behavior of the coronal RPLs, particularly from year 6 to the end of the solar cycle, to some extent reflects the 22-year modulation reported by Badalyan \& Obridko (2017).

\begin{figure*}[h!]
\begin{center}
\includegraphics[width=0.8\textwidth]{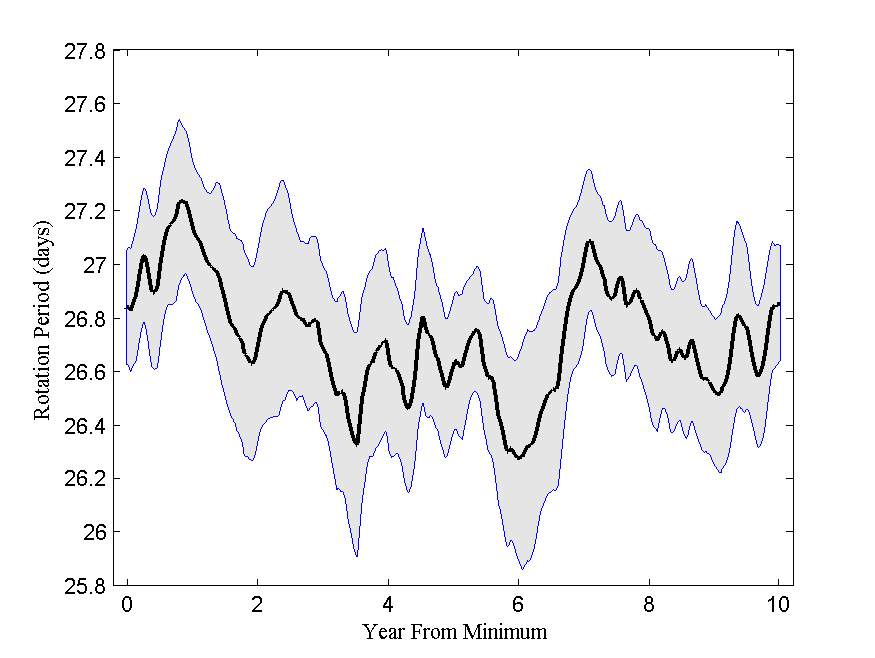}
\end{center}
\caption{Dependence of the coronal RPLs on the solar cycle phase, defined relative to the nearest preceding sunspot minimum. The blue lines indicate the corresponding standard errors.}
\end{figure*}

\subsection{ Periodicity in the Temporal Variation of the Coronal RPLs}
Whether based on tracer tracking or flux modulation, analyzing the periodicity of the derived RPLs in studies of solar rotation helps to reveal the evolutionary characteristics of the rotation and the underlying physical causes (Braj{\v{s}}a et al. 2006; Javaraiah et al. 2009; Javaraiah \& Bertello 2016; Xiang et al. 2026). Thus, we analyzed the periodicity of the coronal RPLs using the CWT method again. The global power spectrum and the 99\% confidence level are shown in Figure 5, where the blue solid line and the black dashed line represent the global power spectrum and the 99\% confidence level, respectively. As shown in Figure 5, the coronal RPLs exhibit periods of approximately 2.5, 6.9, 10.6, 14.4, 19.2, and 32.3 years, all of which exceed the 99\% confidence level. The 2.5 yr period is probably linked to quasi-biennial oscillations (QBOs), and the 10.6 yr period  aligns with the approximately 11 yr solar cycle. Chandra \& Vats (2011) reported that the variation in the coronal rotation period contains a component of approximately 22 years, which is associated with the solar magnetic reversal cycle (Hale's cycle). In the lower solar atmosphere, Ru\v{z}djak et al. (2023) studied the temporal variation of solar rotation derived from the SN data and found a periodicity of about 20 yr, while Braj\v{s}a et al. (2006) also reported a 22 yr periodicity from the investigation of the variation of the sunspot group rotation rate. Thus, in the present study, the 19.2 yr period detected in the coronal RPLs, whose power spectrum displays a broad peak, corresponds to the 22 yr Hale's cycle. The 6.9 yr period (22/3$\approx$7.3 yr) can be regarded as the 1/3 harmonic of the approximately 22 yr Hale's cycle. The 14.4 yr period is nearly consistent with the 14 yr periodicity of sunspot rotation variations reported by Ru\v{z}djak et al. (2023), who attributed this period to the Schwabe cycle. The 32.3 yr period in the coronal RPLs may correspond to the approximately 33 yr periodicity of sunspot rotation variations (Braj\v{s}a et al. 2006; Ru\v{z}djak et al. 2023), and is consistent with 3 times harmonics of the 11 yr Schwabe cycle.

\begin{figure*}[h!]
\begin{center}
\includegraphics[width=0.8\textwidth]{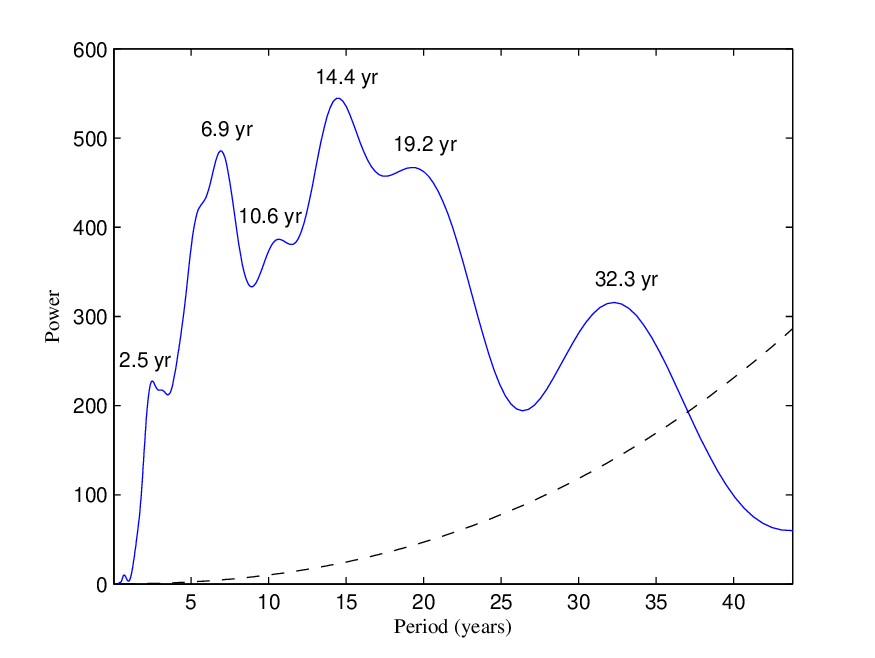}
\end{center}
\caption{Global wavelet power spectrum (solid blue line) and the corresponding 99\% confidence level (dashed black line) for the coronal RPLs during the time interval considered.}
\end{figure*}

\subsection{The XWT and WTC Analysis of the Coronal RPLs and SN}
It is generally accepted that both violent and slowly varying solar phenomena originate from magnetic activities, and that solar activities at different heights in the solar atmosphere correspond to distinct categories (or scales) of magnetic fields (Domingo et al. 2009; De Moortel \& Browning 2015). Moreover, as shown in Figures 3 and 5, the coronal RPLs exhibit multiscale evolution over the considered period of more than 100 years. Given that the XWT and WTC methods are complementary yet distinct in their functions, we combine these two methods to investigate the possible relationship between coronal RPLs and sunspot activity in the time-frequency domain.

\begin{figure*}[h!]
\begin{center}
\includegraphics[width=1.0\textwidth]{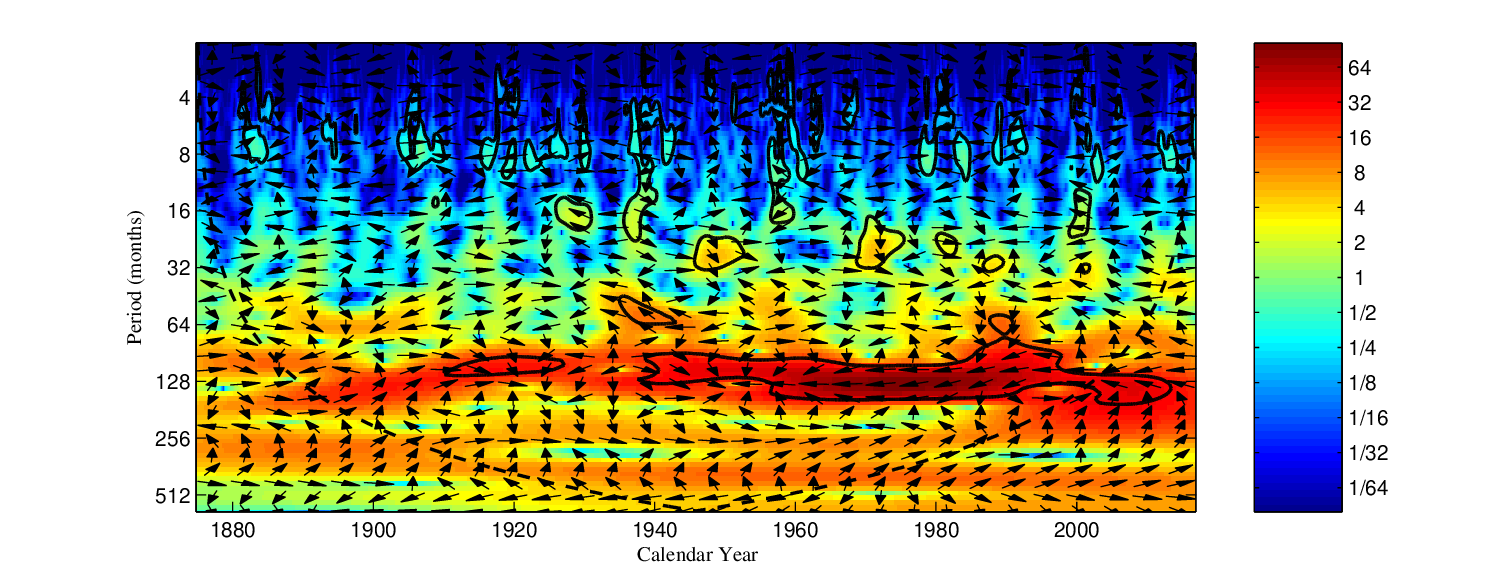}
\includegraphics[width=1.0\textwidth]{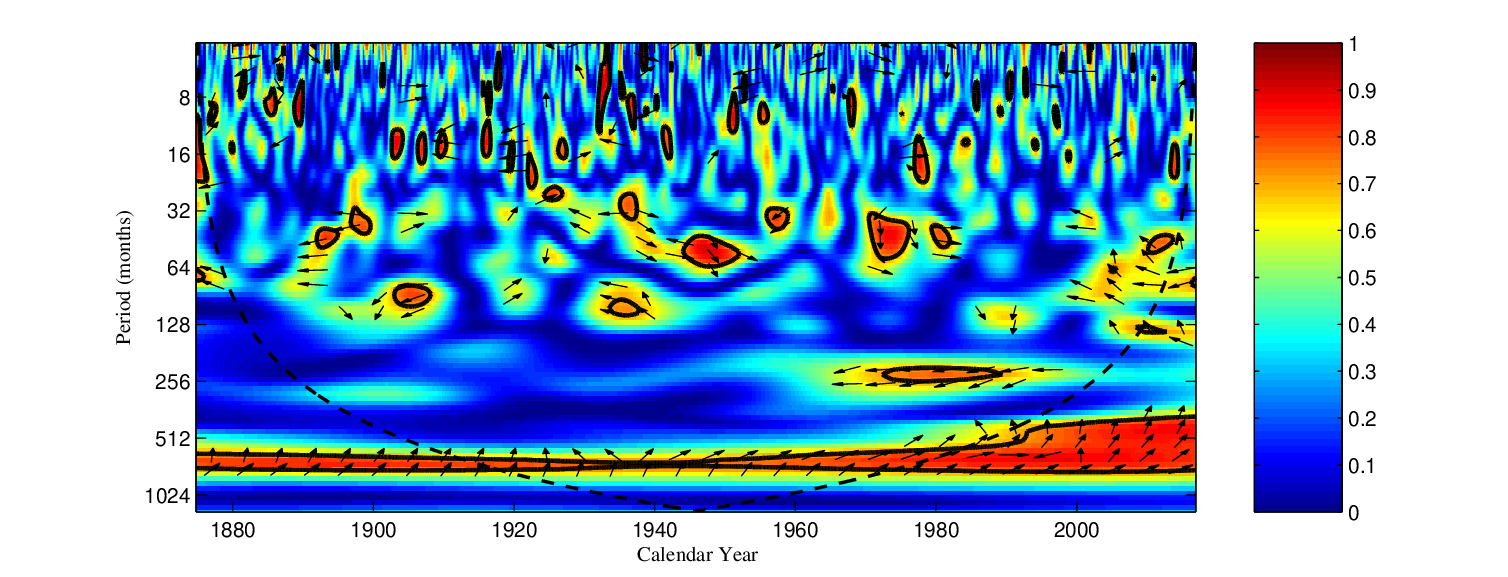}
\end{center}
\caption{The XWT (top panel) and WTC (bottom panel) power spectra of the monthly RPLs and SN over the considered time interval. In each panel, the 95\% confidence level is indicated by thick contours, and the black dashed line marks the cone of influence, beyond which edge effects may affect the results. Arrows denote the relative phase relationship: rightward arrows indicate in-phase behavior, leftward arrows indicate anti-phase, and downward arrows signify that RPLs lead SN by 90$^{\circ}$}
\end{figure*}

The RPLs time series of the corona spans more than 140 years. Such a long duration may cause the power spectra in the XWT and WTC analyses to become densely packed and difficult to resolve. Meanwhile, the RPLs themselves are derived by applying a one-year smoothing to the daily rotation periods obtained from the continuous wavelet power spectrum, which inherently suppresses short-term fluctuations. To clearly analyze the relationship between RPLs and SN in the time-frequency domain over a centennial timescale and within the decadal frequency range, we first computed the monthly means of both datasets and then performed XWT and WTC analyses on the two time series. The XWT and WTC power spectra are shown in the top and bottom panels of Figure 6, respectively.

As shown in Figure 6, on short timescales ranging from approximately 1 to 7 years, the XWT power spectra of RPLs and SN exhibit high-power regions that are discrete, intermittent, small in area, and above the 95\% confidence level. These regions are predominantly concentrated on the QBOs timescale, indicating that the two time series have common periodic oscillations within this frequency (period) range during the corresponding time intervals. Similarly, the WTC power spectra of RPLs and SN also show discrete, intermittent, small-area  high-power regions above the 95\% confidence level, suggesting a correlation between the two time series on this timescale during the corresponding time intervals, although sometimes with relatively low correlation. Combining the XWT and WTC power spectra results, it is evident that on short timescales of about 1 to 7 years, the statistically significant common power between RPLs and SN is mainly concentrated on the QBOs timescale. Therefore, it can be inferred that the variations in RPLs on the QBOs timescale are associated with SN. On the timescale of the approximately 11 yr solar cycle, the XWT power spectra reveal continuous statistically significant high-power regions between RPLs and SN. This can be attributed to the fact that the most typical period of SN is the 11 yr solar cycle, and the preceding analysis also indicates that RPLs exhibit a statistically significant period of 10.6 yr. However, on this timescale, the WTC power spectra of RPLs and SN do not show statistically significant high-power regions, suggesting a weak correlation between the variations of the two variables on this timescale. Combining the XWT and WTC results, it can be inferred that although the RPLs exhibit a periodic variation of about 11 years, their variations on this timescale are weakly correlated with SN. On timescales of several decades, the high-power regions in the XWT power spectra lie outside the cone of influence, where edge effects may introduce uncertainties; therefore, they are not presented in the spectra. On the timescale of approximately 50 to 65 yr, however, the WTC power spectra of RPLs and SN display a continuous statistically significant high-power region spanning the entire time interval considered. Although part of this region falls outside the cone of influence, it still suggests a correlation between RPLs and SN on this timescale. Moreover, most of the arrows within this region point upward, indicating that SN variations lead those of RPLs. This result corresponds to the long-term trends of RPLs and SN shown in Figure 3: both series exhibit turning points within this timescale, and the long-term trend of SN appears to drive that of RPLs.  On longer timescales, the WTC power spectra, similarly to the XWT spectra, fall outside the cone of influence, where both the statistical significance of the common power and the phase estimates between RPLs and SN become unreliable due to edge effects and the finite length of the time series. We therefore do not extend our interpretation to these timescales.

\section {Discussions and Conclusion}
The 10.7 cm radio flux time series from 1874 May 9 to 2017 June 30, which was reconstructed by Yeo, Solanki, \& Krivova (2020) and represents the longest available observational proxy of the solar corona, is used to extract coronal RPLs using the CWT. Subsequently, we apply linear fitting and wavelet analysis methods, such as XWT and WTC, to examine the multi-scale evolution of coronal rotation over the time interval of more than 140 years and to explore the underlying physical mechanisms of these evolutionary patterns. Additionally, the 10.7 cm radio flux represents the radio emission intensity across the entire solar disk and is not differentiated by latitude or longitude. Accordingly, the coronal PRLs examined in this study are derived from a global perspective of the Sun, are independent of latitudinal differential rotation, and can be considered as averages over global latitudes. This global perspective offers distinct advantages in the study of coronal rotation, as it avoids interference from fine-scale structures, localized distributions, and short-term evolutionary features (Xu et al. 2020; Singh et al. 2021), enabling the tracking of coronal rotation periods with high temporal resolution and revealing multi-scale variations.

The CWT analysis of the 10.7 cm radio flux indicates that the coronal RPLs exhibit complex temporal evolution over the 143-year time interval considered. The dominant synodic rotation period is 26.72 days, corresponding to a sidereal rotation period of 24.90 days, which is consistent, within the margin of error, with the average sidereal rotation period of 24.3 days obtained by Chandra \& Vats (2011) from the 10.7 cm radio flux for the period 1947-2009. Furthermore, this coronal rotation period is shorter than the typical synodic period of approximately 27.4 days associated with large-scale magnetic activity in the underlying photosphere (Balthasar 2007; Li et al. 2019), indicating that the corona rotates faster than the lower photosphere. Although this finding is consistent with several previous studies (Vats et al. 2001; Li et al. 2019; Singh et al. 2021; Sharma et al. 2020, 2023; Xiang et al. 2023), the present study, spanning over 140 years, confirms the faster coronal rotation for the first time over such an extended timescale, thereby increasing the robustness of the conclusion to some extent. In the corona, the energy density of the plasma is lower than that of the magnetic field, resulting in a regime where the magnetic field dominates and the frozen-in effect is strong. In this regime, the magnetic field topology largely constrains the coronal plasma motions. In contrast, the opposite situation occurs in the photosphere, where the magnetic field is dragged by the plasma motion (Wiegelmann et al. 2014; Bellot Rubio \& Orozco Su{\'a}rez 2019; Alissandrakis 2020). Moreover, the large-scale magnetic structures rotate more slowly than small-scale magnetic concentrations, but faster than the photospheric plasma (Komm et al. 1993; Xiang et al. 2014; Xu \& Gao 2016; L{\"o}{\ss}nitz et al. 2025). Owing to the fact that the ubiquitous small-scale magnetic fields can effectively drive the strongly frozen-in corona (Li et al. 2019; Xiang et al. 2023; Rao et al. 2024), the entire coronal plasma atmosphere rotates at a faster rate than both the underlying photospheric plasma and the large-scale magnetic structures.

Over the past more than 100 years, the fitting results in Figure 3 show that the relationship between the coronal RPLs and SN changes over time: anticorrelation during 1874.9-1943.1, correlation during 1943.1-1958.2, anticorrelation during 1958.2-1987.4, and correlation during 1987.4-2016.9. Overall, the RPLs remain predominantly anticorrelated with the SN, i.e., rotation becomes faster as activity increases, except for the two intervals 1943.1-1958.2 and 1987.4-2016.9, in which a correlation is observed. Over the entire time interval considered, the correlation coefficient between the long-term trend fits of the coronal RPLs and those of SN is -0.472. Its absolute value is much greater than the critical value of 0.014 at the 99.9\% confidence level for their correlation. If only the approximately 97 years during which they exhibit an anticorrelation are considered, the correlation coefficient is -0.783, indicating a significant negative correlation; its absolute value is also much greater than the critical value of 0.017 at the 99.9\% confidence level for their correlation. Therefore, the overall negative correlation between the long-term variations of the coronal RPLs and SN is statistically significant.
However, the long-term variation of the coronal RPLs does not always follow an inverse trend relative to SN; instead, the two sometimes exhibit the same trend, particularly during the recent decades spanning 1987.4-2016.9. This same trend is consistent with the results reported by Li et al. (2012) and Xie et al. (2017), which were also derived from 10.7 cm radio flux data. In recent years, several studies have consistently revealed that the corona, which rotates faster than the underlying photosphere, is primarily driven by the ubiquitous small-scale magnetic fields on the solar surface (Li et al. 2019; Xiang et al. 2023; Rao et al. 2024). Among them, Xiang et al. (2023) further demonstrated that the stronger the solar activity, the stronger the control of small-scale magnetic fields over coronal rotation. This control mechanism exactly explains the gradually decreasing trend in RPLs from 1874.9 to 1943.1. During this time interval, as shown in top panel of Figure 3, the solar magnetic activity (SN) gradually increases, corresponding to a gradually stronger control effect of the faster-rotating small-scale magnetic fields on coronal rotation, and consequently the coronal rotation gradually accelerates.

On the other hand, based on sunspot, sunspot group, and global magnetic field data obtained from different periods, multiple studies have consistently shown that the rotation rate of the solar photosphere is higher during time intervals of low magnetic activity than during those of high magnetic activity (Hathaway \& Wilson 1990; Braj{\v{s}}a et al. 2006; Obridko \& Shelting 2016). This phenomenon arises because larger sunspot groups exhibit slower rotational rates compared to their smaller counterparts, and strong magnetic fields generally suppress rotation on the solar surface (Howard 1984; Xu \& Gao 2016; Li et al. 2019). When more large-scale magnetic activity regions (such as sunspot groups) appear on the photosphere after originating below, the rotation of the underlying photosphere slows down, and vice versa. This deceleration effect in the photosphere gradually propagates upward, influencing the solar atmosphere above the photosphere and ultimately reaching the corona. As shown in the top panel of Figure 3, the SN exhibits a gradually increasing trend from 1874.9 to 1958.2, indicating a progressive enhancement of solar magnetic activity. This enhanced magnetic activity gradually reduces the rotation rate of the solar photosphere, thereby affecting the coronal rotation. When this deceleration effect outweighs the acceleration trend of coronal rotation driven by small-scale magnetic activity, the coronal rotation trend shifts from acceleration to deceleration. This result is illustrated in the middle and bottom panels of Figure 3: during 1943.1-1958.2, the SN shows a gradually increasing trend, while the coronal rotation exhibits a gradual deceleration.

The  explanation for the long-term variation trend of coronal rotation from 1874.9 to 1958.2 is based on the physical  basis of the gradually increasing SN. In contrast, during 1958.2 to 2016.9, the long-term trend of SN shifts to the gradual decrease, and accordingly, the influence of the solar magnetic activity level on the rotation of the solar atmosphere, as well as the modulation of coronal rotation by small-scale magnetic fields, should operate in the opposite direction. This reversed mechanism precisely accounts for the changes in the long-term trend of coronal rotation from 1958.2 to 2016.9: coronal rotation exhibits the gradual deceleration from 1958.2 to 1987.4, and the gradual acceleration from 1987.4 to 2016.9. In summary, the long-term variation trend of the coronal RPLs results from the combined effect of two factors: the dominant control of the faster-rotating ubiquitous small-scale magnetic fields over coronal rotation, and the modulation of the solar magnetic activity level (SN) on the rotation of the entire solar atmosphere.

As indicated by the dependence of the coronal RPLs on the solar cycle phase, compared with the results about the dependence of the underlying photospheric strong magnetic activity (SN) on the solar cycle phase (Li et al. 2013; Xiang et al. 2014), the coronal rotation is faster than the rotation of the underlying large-scale magnetic fields at all phases of the solar cycle. Specifically, during periods of weak solar activity, the RPLs show larger values, i.e., the coronal rotation is slower; whereas during periods of strong solar activity (approximately from the 3rd to the 6th year of the solar cycle), the coronal rotation is faster. This result is opposite to the average rotation characteristics of photospheric sunspots; the latter rotates faster during the solar minimum than during the high active phase (Hathaway \& Wilson 1990; Braj{\v{s}}a et al. 2006), but is in general consistent with the findings of Mouradian et al. (2002) and Xie et al. (2017), both of which indicated that the coronal RPLs are relatively longer during the solar minimum. Therefore, from the phase relationship of the coronal RPLs over the solar cycle, it can be inferred that the coronal rotation is only weakly related to the SN, and its faster rotation is precisely consistent with previous results showing that coronal rotation is mainly modulated by fast-rotating small-scale magnetic fields (Li et al. 2019; Xiang et al. 2023; Rao et al. 2024). Moreover, the stronger the solar activity, the stronger the modulation of coronal rotation by small-scale magnetic activity (Xiang et al. 2023). Correspondingly, from the perspective of the solar cycle phase, during periods of strong solar activity, the enhanced control of the faster-rotating small-scale magnetic activity corresponds to the faster coronal rotation; whereas during periods of weak solar activity, the weakened modulation of coronal rotation by small-scale magnetic fields corresponds to the slower coronal rotation. This scenario explains the pattern of the dependence of the coronal RPLs on the solar cycle phase as shown in Figure 4. In detail, this figure also shows that the RPLs exhibit a three-peak fluctuation during the solar maximum. The timescale of this fluctuation is comparable to that of the QBOs, suggesting that these fluctuations are related to the QBOs of sunspot activity. Essentially, this may stem from variations in the magnetic activity level induced by the QBOs of sunspots, which in turn affect the modulation of coronal rotation by small-scale magnetic activity.

Several studies of coronal differential rotation have also examined how coronal rotation varies with the solar activity cycle (Sime et al. 1989; Altrock  2003; Tlatov 2006; Badalyan \& Obridko 2018; Edwards et al. 2022). For instance,  Sime et al. (1989), Altrock (2003) and Edwards et al. (2022) reported that slower coronal rotation was observed in the early stage of the ascending phase. This is consistent with our results shown in Figure 4, in which RPLs increase rapidly from the beginning of the solar cycle to its first year. However, Sime et al. (1989) considered temporal variations in coronal rotation to be unrelated to the solar activity cycle, and most of these variations appeared to be associated with higher latitudes. Altrock (2003) identified pronounced solar cycle variations in coronal rotation, with high- and low-latitude rotational changes occurring nearly in antiphase. At latitudes below approximately  $60^\circ$, the rotation is nearly rigid near solar minimum and becomes more differential during the ascending and maximum phases. At latitudes above approximately $60^\circ$, the pattern is reversed: rotation remains nearly rigid during the ascending and maximum phases but becomes strongly differential around solar minimum, almost opposite in phase to the low-latitude behavior.  Tlatov (2006) confirmed this antiphase behavior and further demonstrated that high-latitude rotational variations generally exhibit larger amplitudes. In particular, rotation at low and middle latitudes decelerates near solar activity maxima, whereas high-latitude rotation decelerates near solar activity minima. To some extent, these high-latitude rotational variations are consistent with our results. Moreover, in terms of the latitudinal dependence of the rotation period with respect to the phase of the solar activity cycle, Badalyan \& Obridko (2018) reached conclusions consistent with those of Tlatov (2006): low-latitude (equatorial) rotation is slowest, i.e., exhibits the longest period, near solar maxima, whereas high-latitude rotation is slowest near solar minima. Because our analysis relies on disk-integrated 10.7 cm radio flux, the RPLs derived in this study are expected to represent the global solar rotation. The complex temporal variations in the RPLs may be influenced by rotation at both high and low latitudes. The deceleration of global rotation around solar activity minima could arise from the superposition of time-dependent variations in rotation at high and low latitudes.

Given the multi-scale temporal variations exhibited by the coronal RPLs, we employed the CWT to analyze their periodicity and identified statistically significant periods at 2.5, 6.9, 10.6, 14.4, 19.2, and 32.3 years. Among these, the 2.5-year period is attributed to the QBOs. The QBOs represent the most significant periodicities detected in observational data of solar activity and its indicators, aside from the rotation period and the solar cycle. They are thought to originate at the base of the solar convection zone, with their influence extending from the solar interior to the heliosphere (Bazilevskaya et al. 2000, 2014; Vecchio et al. 2012; Oloketuyi et al. 2024). In the RPLs across different layers of the Sun, QBOs have been detected from the underlying photosphere to interplanetary space. For instance, periods of approximately 2.6 and 3.1 yr are found in sunspots (Javaraiah \& Gokhale 1995; Li et al. 2011), the period of about 3.2 yr in the chromosphere (Xiang et al. 2026), and periods of about 2.9 and 3.4 yr in the interplanetary magnetic field (Xiang et al. 2020). The 6.9 yr period may be the 1/3 harmonic of the approximately 22 yr Hale's cycle. However, as noted by Ru\v{z}djak et al. (2023), the analysis of the spectral window demonstrated that periods in the range of 2.5 to 10 years are unreliable in both solar activity and solar rotation. This is because they are influenced by aliasing caused by the Schwabe cycle, as well as by long-term variations and harmonics of quasi-annual and quasi-biennial periods. Therefore, the 6.9 yr period, and possibly even the 2.5 yr period in the coronal RPLs, may be affected by Schwabe cycle aliasing and by annual and biennial harmonics. The two period values obtained here should therefore be regarded only as averages over the time span of more than a hundred years, influenced by these factors.

For the period of approximately 11 yr in the RPLs, it has been detected in various layers of the solar atmosphere (Obridko \& Shelting 2001; Braj{\v{s}}a et al.2006; Javaraiah \& Bertello 2016; Sharma et al. 2024; Xiang et al. 2026). Moreover, the 14 yr period obtained from the study of the rotation variation of SN in the lower solar atmosphere was considered to be related to the Schwabe cycle (Ru\v{z}djak et al. 2023). Consequently, the 10.6 yr and 14.4 yr periods identified in the temporal evolution of the coronal RPLs should correspond to 11 yr Schwabe cycle. However, as shown in Figure 5, the signal of the 10.6 yr period is relatively weak, mainly because the coronal rotation is weakly correlated with the large-scale magnetic activity in the underlying photosphere, i.e., the modulation effect of the latter on coronal rotation is weak (Li et al. 2019; Xiang et al. 2023; Rao et al. 2024). Due to this weak signal, Chandra \& Vats (2011), using 10.7 cm radio flux data over the period 1947-2009, did not detect the 11 yr solar cycle in their study of coronal rotation, but instead identified the 22 yr solar magnetic reversal cycle. Based on the modified coronal index during 1939 to 2019, Deng et al. (2020) also suggested the temporal variations of solar rotation are  connected with the 22 yr solar magnetic reversal cycle. In addition, Tlatov (2006) demonstrated that the rotational velocity residuals at low ($\pm$ $10^{\circ})$ and high $(60^{\circ}$-$70^{\circ})$ latitudes vary with time and exhibit a clear modulation of approximately 22 yr. In the lower atmosphere, the period of about 20 yr was detected in the temporal variation of solar rotation derived in SN  by Ru\v{z}djak et al. (2023), and the 22 yr period from the investigation of the variation of the sunspot group rotation rate was reported in Braj\v{s}a et al. (2006). Similarly, Poljan\v{c}i\'c Beljan et al. (2022), using sunspot group data from the Kanzelh\"ohe Observatory for Solar and Environmental Research for the period 1964--2016, found a 22-yr periodicity in the equatorial rotation rate. Taking all these results into consideration, the 19.2 yr period, corresponding to the broad peak in the power spectrum of the coronal RPLs, is closely related to the 22 yr solar magnetic reversal cycle. The 32.3 yr period of the coronal RPLs has scarcely been discussed in previous studies of coronal rotation variations. Howerver, similar periods have been discussed in detail in studies of solar rotation based on sunspot data. For instance, the $35\pm3$ yr period in sunspot rotation variations (Ru\v{z}djak et al. 2023) and the approximately 33 yr period in sunspot group rotation variations (Braj\v{s}a et al. 2006) were both considered to be dominated by the 11 yr Schwabe cycle. Therefore, the 32.3 yr period in the coronal RPLs should also correspond to the 11 yr Schwabe cycle and can be regarded as the 3 times harmonics of the 11 yr Schwabe cycle.

Finally, by combining the XWT and WTC methods, we analyzed the time-frequency relationship between the coronal RPLs and the SN, further confirming the aforementioned findings. On the QBOs timescale, the certain degree of correlation is observed between the coronal RPLs and the SN. On the timescales of the solar cycle, although both time series exhibit the solar cycle signal, they are not correlated with each other.  On longer timescales of approximately 50-65 years, however, the significant correlation appears between the two series in the WTC analysis, indicating that the long-term variation of SN may influence the long-term trend of the coronal RPLs. Gleissberg (1939) noted that the amplitudes of solar activity cycles between 1750 and 1928 exhibit a periodicity of 7-8 solar cycles. Subsequent studies have suggested that the Gleissberg cycle is not strictly periodic but may consist of two distinct components: a period of 90-100 years and another of 50-60 years (Rozelot 1994; Hathaway et al. 1999; Ogurtsov et al., 2002; Hathaway 2015), with the latter possibly representing the first subharmonic of the Gleissberg cycle (Tan 2011; Luo \& Tan 2024). Furthermore, Javaraiah (2003) found that the equatorial rotation rate  derived from sunspot groups decreased significantly from cycle 13 to 14 and again from cycle 21 to 22, with a time gap of about 90 years between these two drops, which suggests a connection to the Gleissberg cycle. Thus, the significant correlation between the coronal RPLs and SN on the timescales of 50 to 65 years suggests that the long-term evolution of coronal RPLs may be associated with the Gleissberg cycle.  Hathaway (2015) identified that the preceding minimum and maximum of the Gleissberg cycle occurred around solar cycles 14 and 19, respectively. The results shown in Figures 3 and 6 reveal that the long-term trend of coronal RPLs does not directly coincide with these Gleissberg extrema; rather, the transitions in the RPL trends occur after the corresponding Gleissberg maxima or minima, indicating that the Gleissberg cycle appears to lead the long-term variations of the coronal RPLs.
This lag is consistent with the phase shift between solar activity and the equatorial rotation rate, as reported by Poljan\v{c}i\'c Beljan et al. (2022), further supporting the inference that the Gleissberg cycle leads the long-term variations in solar rotation. In essence, this is due to the modulation of the rotation of the entire solar atmosphere by the long-term variation of the solar magnetic activity level over extended timescales.

In summary, the coronal RPLs exhibit the complex evolutionary pattern, which fundamentally results from the combined modulation of two factors: the direct and dominant effect of the ubiquitous small-scale magnetic fields on the coronal rotation, and the modulation of the entire solar atmospheric rotation by the solar magnetic activity level.

\begin{acknowledgements}
\section*{Acknowledgements}
The authors acknowledge with gratitude Yeo, Solanki, \& Krivova for reconstructing the 10.7 cm radio flux data.  This work is supported by the National Key R \& D Program of China (grant No. 2025YFF0510701), the National Natural Science Foundation of China (Grant Nos. 12673064, 42530206, 12373059,  42474224), the Innovation Team Project of Yunnan Revitalization Talent Support Program (202405AS350012), the Beijing Natural Science Foundation (1242035), the Climbing Program of NSSC (E6PD3001), the Project Supported by the Specialized Research Fund for State Key Laboratories, and the Chinese Academy of Sciences.
\end{acknowledgements}

\end{document}